# Evaluation of Tropical Cyclone Track and Intensity Forecasts from Artificial Intelligence Weather Prediction (AIWP) Models

Mark DeMaria[a], James L. Franklin[b], Galina Chirokova[a], Jacob Radford[a],

Robert DeMaria[a], Kate D. Musgrave[a] and Imme Ebert-Uphoff[a,c]

[a] *Cooperative Institute for Research in Atmosphere, Colorado State University, Fort Collins, CO*

[b] *Lynker, Miami, FL*

[c] *Electrical and Computer Engineering, Colorado State University, Fort Collins, CO*

*Corresponding author*: Mark DeMaria, Mark.DeMaria@colostate.edu

ABSTRACT

In just the past few years multiple data-driven Artificial Intelligence Weather Prediction (AIWP) models have been developed, with new versions appearing almost monthly. Given this rapid development, the applicability of these models to operational forecasting has yet to be adequately explored and documented. To assess their utility for operational tropical cyclone (TC) forecasting, the NHC verification procedure is used to evaluate seven-day track and intensity predictions for northern hemisphere TCs from May-November 2023. Four open-source AIWP models are considered (FourCastNetv1, FourCastNetv2-small, GraphCast-operational and Pangu-Weather).

The AIWP track forecast errors and detection rates are comparable to those from the best-performing operational forecast models. However, the AIWP intensity forecast errors are larger than those of even the simplest intensity forecasts based on climatology and persistence. The AIWP models almost always reduce the TC intensity, especially within the first 24 h of the forecast, resulting in a substantial low bias.

The contribution of the AIWP models to the NHC model consensus was also evaluated. The consensus track errors are reduced by up to 11% at the longer time periods. The five-day NHC official track forecasts have improved by about 2% per year since 2001, so this represents more than a five-year gain in accuracy. Despite substantial negative intensity biases, the AIWP models have a neutral impact on the intensity consensus. These results show that the current formulation of the AIWP models have promise for operational TC track forecasts, but improved bias corrections or model reformulations will be needed for accurate intensity forecasts.

SIGNIFICANCE STATEMENT

Weather prediction (WP) models based on artificial intelligence (AI) have proliferated over just the past few years. Once trained, these models run very quickly, making them appealing for operational forecasting. This study is one of the first to consistently evaluate TC track and intensity forecasts from four open-source AIWP models using NHC verification procedures to determine their applicability for operational forecasting. The track forecast errors are comparable to those from the best operational models. However, the intensity forecasts have no skill relative to the simplest statistical models, due to a substantial low bias. Thus, the TC track forecasts can be used directly as guidance, but post-processing bias corrections or AIWP model modifications will be needed for useful TC intensity forecasts.

# 1. Introduction

In just the past few years, many data-driven global weather prediction models based on artificial intelligence have been developed. One of the first of these models, referred to as Artificial Intelligence Weather Prediction (AIWP) models, was FourCastNetv1 (hereafter FCN1), developed by the NVIDIA corporation in 2022 (Pathek et al. 2022). This model was trained on ERA5 0.25º reanalysis fields (Hersbach and co-authors 2020) using a sophisticated machine learning method, namely a Fourier-based neural network model. The model is entirely empirical and does not directly include physically based equations. Although the model requires considerable computational resources to train, predictions with the trained model run very quickly. For example, an implementation of FCN1 on a mid-range server with one GPU (NVIDIA, RTX A6000) produces a ten-day global forecast in about a minute.

Following the release of FCN1, many other AIWP models became available, including an updated version of FCN1 referred to as FourCastNetv2-small (Bonev et al. 2023, hereafter FCN2), Pangu-Weather (Bi et al. 2023, hereafter PGWX), and GraphCast-operational (Lam et al. 2023, hereafter GRCS). The ECMWF is developing its own AIWP model (Lang et al. 2023), and new versions of these and other models are becoming available almost monthly. AI methods are also being used to develop ocean models (Chattopadhyay et al. 2023).

As one of the first models of its type, FCN1 surprised the community in terms of potential, but also had several shortcomings. It tended to become unstable at longer lead times, with instability typically starting at the poles and spreading equatorward. This issue was addressed in FCN2 (Bonev et al. 2023) using methods customized for spherical coordinates. Despite those known issues for FCN1, we include FCN1 in this analysis to gain insight into the evolution of AIWP models and the effect of their shortcomings on tropical cyclone (TC) track and intensity forecasting. The instability noted above was seen in only a small fraction (< 1%) of the cases in the regions around the TCs examined in this study.

Given the rapid proliferation of AIWP models, it will take some time to perform comprehensive evaluations and determine the models' utility for operational forecasting (Ebert-Uphoff and Hilburn 2023). Preliminary studies have shown that the AIWP model errors are comparable to or better than traditional physics-based models in terms of global 500-hPa geopotential height anomalies and other large-scale metrics (e.g., Lam et al. 2023, Lang et al. 2023). The very low computational cost of AIWP models, once trained, also holds great promise for generating very large ensembles to provide improved guidance for forecast uncertainty products. However, there are indications that the spread of AIWP ensembles is highly dependent on the scale of the weather forecast phenomena of interest, such that the larger-scale weather systems show good spread, while smaller scale disturbances do not (Selz and Craig 2023). Thus, more research is needed to determine if the potential of large ensembles can be realized in practice.

Several studies have evaluated the errors in AIWP TC track forecasts. For example, Lam et al. (2023) showed that GRCS had lower median track errors than the deterministic ECMWF model for a large sample of cases from 2018-2021, with improvements of about 10% at day 5. This is a very encouraging result because the ECMWF has usually been among the best-performing track models over the past several years (e.g., Cangialosi et al. 2023). Verifications of FCN1 TC track forecasts suggested that that model also had errors comparable to some of the best deterministic track forecasts (Pathek et al. 2022).

A limitation of the previous evaluations is that they used varying trackers and non-standard methodologies. Thus, it is difficult to use those results to draw conclusions about how these AIWP models might contribute to operational forecasting. For example, Xie et al. (2024) limited their verification sample to only stronger TCs that made landfall in China. As a first step towards evaluation of AIWP for U.S. operational needs, the TC track and intensity predictions from four AIWP models (FCN1, FCN2, PGWX, and GRCS) are verified following procedures used by NHC to evaluate their own official forecasts and operational guidance. These four models were chosen because they are open-source, allowing the Cooperative Institute for Research in the Atmosphere (CIRA) to run the AIWP models in near real-time and to produce a database of retrospective runs. Despite being data-driven, the AIWP models still require an initial condition. For the consistency of the CIRA model database, the publicly available GFS analysis was used for initialization of the four AIWP models.

In this study, a common tracker developed at CIRA (hereafter, the CIRA TC tracker) is applied to the four AIWP models to facilitate intercomparisons. The standard NHC verification metrics, including the mean track error, mean absolute intensity error, and mean intensity bias, were calculated using the NHC verification software package. A new metric, detection rate, was added to the NHC verification software; this measures the ability of a model to retain a trackable circulation. To help determine the applicability of AIWP models to operational TC forecasting, the results are compared to those from representative operational track and intensity models.

NHC's best performing track and intensity models are "consensus" models, combinations of typically high-performing individual models. Sometimes, models that do not rank among the best performers still produce a positive contribution when added to the consensus if they provide independent information (Cangialosi et al. 2023, Simon et al. 2018). Thus, in this study we also test the impact of adding AIWP forecasts to the operational consensus. To get a feel for the applicability of AIWP models to ensemble forecasting, the spreads of the four AIWP track and intensity forecasts are calculated and compared with those from four representative operational models.

Section 2 summarizes the AIWP models, section 3 describes the CIRA TC tracker, section 4 describes the verification methodology, section 5 shows the verification results, and section 6 presents the consensus and ensemble-spread results. Concluding remarks are presented in section 7.

## 2. Input data

*a. AIWP model fields*

Open-source versions of the four AIWP models were implemented at CIRA and run to 10 days using initial conditions from the NCEP/GFS model (NCEP 2024). Some of the models, such as PGWX, have multiple versions that use different time steps. For the CIRA archive, the versions of the models with a 6-h time step were used. GRCS also has options for the number of vertical levels; the 13-level version was implemented at CIRA. The output was converted to a compressed NetCDF format and archived on a public-facing website (see data availability statement). The forecast fields were saved every 6 h on a 0.25º lat/lon grid. The models were initialized every 12 h from Oct 2020 to Apr 2023 and every 6 h from May to Nov 2023. To allow a comparison with the most recent operational model forecasts that include four runs per day (00, 06, 12 and 18 UTC), only the May-Nov 2023 cases were included in this study.

*b. Tropical cyclone data*

TC data were obtained from archives of the Automated Tropical Cyclone Forecast (ATCF) system (Sampson and Schrader 2000). The ATCF includes real-time and forecast estimates of the position and intensity (maximum 1–min-mean surface winds) of all TCs occurring across six global basins: the north Atlantic Ocean (AL), the eastern north Pacific Ocean to 140°W (EP), the central north Pacific Ocean from 140°W to the Dateline (CP), the western North Pacific Ocean west of the Dateline (WP), the north Indian Ocean (IO), and the south Indian Ocean and south Pacific region (SH). NHC has forecast responsibility for the AL and EP basins, the Central Pacific Hurricane Center (CPHC) has responsibility for the CP basin, and the Joint Typhoon Warning Center (JTWC) has U.S. responsibility for the remaining basins (WP, IO, and SH). Each agency maintains its own archive of ATCF files.

This study used the "a-decks" and "b-decks" from the ATCF. A-decks include operational initial positions and intensities for each TC, and forecasts from operational numerical models as well as the NHC, CPHC, and JTWC official forecasts. Storm structure information (wind radii) is also available in the a-decks, but those data were not evaluated in this study. The b-decks contain the verifying ground truth positions and intensities used to evaluate the TC forecasts – the so-called "best tracks", after-the-fact estimates of TC track, intensity, and storm structure parameters at 6-hourly synoptic times. The b-decks also includes an assessment of the cyclone type (e.g., tropical, subtropical, extratropical, disturbance, etc.). NHC, CPHC and JTWC are responsible for the best tracks for storms (or portions of storms) in their respective areas of responsibility. The data used here represent the final post-storm best tracks from NHC for the AL and EP basins and from JTWC for the WP and IO basins. Intermediate ("working best tracks") data were used for CP cases because they did not finalize their best tracks by the time of this study. The individual TCs in the ATCF database are referenced by a storm name (e.g., Idalia) and an identification label comprised of the basin, storm number and year (e.g., AL102023 for the 10$^{th}$ TC in the Atlantic basin in 2023).

Table 1 lists the forecast models used in the current study and their ATCF identifiers. For most of the models, two ATCF IDs are included; these correspond to the "late" and "early"

File generated with AMS Word template 2.0

version of each model, respectively. Late versions represent track and intensity forecasts extracted directly from dynamical model output, but in most cases these dynamical models are not available in time to meet operational forecast deadlines. Therefore an "early" version of the model is created for forecasters by adjusting the most-recently-available late-model forecast (usually 6- or 12-h old) to match the TC's current initial conditions. This process is described in greater detail in section 4 and was also applied to the AIWP models. Further details on the operational models in Table 1 can be found in DeMaria et al. (2022) and Cangialosi et al. (2023).

| Forecast Model/Method | ATCF Identifier | Forecast/Model Type |
|---|---|---|
| NHC or CPHC official forecast | OFCL | Subjective |
| JTWC official forecast | JTWC | Subjective |
| FourCastNetv1 | FCN1/FC1I | AIWP |
| FourCastNetv2-small | FCN2/FC2I | AIWP |
| GraphCast-operational | GRCS/GRCI | AIWP |
| Pangu-Weather | PGWX/PGWI | AIWP |
| NCEP GFS | GFSO/GFSI | Global NWP |
| GFS with simplified CIRA TC tracker | GFSS (late) | Global NWP |
| U.K. Met Office global model | UKM/UMKI | Global NWP |

| ECMWF Integrated Forecast System | ECMO[1]/ECMI | Global NWP |
|---|---|---|
| HWRF | HWRF/HWFI | Regional NWP |
| COAMPS-TC | CTCX/CTCI | Regional NWP |
| SHIPS | DSHP | Statistical-dynamical |
| LGEM (Logistic Growth Equation Model) | LGEM | Statistical-dynamical |
| Trajectory CLIPER | TCLP | Statistical |

Table 1. Forecasts used in the verification. For models with two Tech IDs, the first entry is the late model and the second is the early model. Unless otherwise indicated, if only one Tech ID is listed in column 2, it is an early model. All forecasts except those from the AIWP models and GFSS were obtained from ATCF databases.

## 3. The CIRA TC tracker

The purpose of a tracker is to identify the location of the TC center and estimate the intensity from model forecast fields. Operational TC trackers are also used to determine storm structure parameters, such as the radii of 34-, 50- and 64-kt winds in various directions around the TC center, but that application will not be considered here. There are two basic types of trackers. The first type does not require a first-guess position and is used for TC genesis forecasting and in climate model studies (e.g., Jing et al. 2021). The second type is used to find the center and intensity of existing TCs that are already being tracked by operational agencies. The tracker used in this study is of the second type. For this application, the first guess for the tracker is obtained from the operational position estimates or a short-term forecast.

---

[1] JTWC splits ECMWF (ECMWF 2024) forecasts into two ATCF IDs depending on the model initialization time: ECMF is used to designate a 0000 or 1200 UTC forecast, while ECMO refers to the 0600 and 1800 UTC forecasts. NHC and CPHC do not make this distinction, using the ECMO designation regardless of the forecast's initial time. In this manuscript, the ECMO designation represents both ECMF and ECMO forecasts.

Marchok (2021) summarizes the TC trackers used in operational models. Even though it is nominally the surface wind center that is being sought, the most robust trackers use several variables (horizontal wind, geopotential height, sea-level pressure) from multiple model levels. However, some of the AIWP models have limited variables and vertical levels, so a simpler version was required here. Marchok showed that the most accurate single-level trackers for center location were those that used data from the 850-hPa level. That study evaluated two versions of a simple tracker; one that finds the minimum in the 850-hPa geopotential height and one that maximizes the 850-hPa wind circulation. Of the two, the 850-hPa wind circulation method was slightly more effective at tracking TCs longer into the forecast period, and so was chosen for this study. The wind tracker has an additional advantage in that a similar version is used for vortex removal in the NHC operational Statistical Hurricane Intensity Prediction Scheme (SHIPS) (DeMaria et al. 2022), which was adapted for this study. For intensity, the 10-m winds are used since those were available from all four AIWP models. Because the SHIPS TC tracker is different from the version described by Marchok and was further modified for use with the AIWP models, the method used for the CIRA TC tracker is described in more detail in the Appendix.

## 4. Verification methodology

The CIRA TC tracker described in the Appendix was applied to every northern hemisphere TC during the period May-Nov 2023 that NHC, JTWC, or CPHC was monitoring. As the U.S. operational centers do, all cases classified as a TC (depression, storm, and hurricane/typhoon), as well as subtropical cyclone cases, in the best track were included in the verification results without regard to intensity, convective organization, or other selection factors. The extratropical stage was excluded.

The CIRA TC tracker was applied to the AIWP models out to 180 h; this allowed early models to be generated out to 168 h, the current extent of internal official forecasts being prepared by NHC. The 180 h forecasts were needed for the rare cases when a 6-h old late model forecast was not available, and the 12-h forecast was used instead. While AIWP

File generated with AMS Word template 2.0

models run very quickly, they are still considered late models because when run in real time, they would still need to wait for a GFS or ECMWF analysis to become available for initialization. Recall that early versions of the dynamical models are needed so as to have dynamical guidance available in time for operational forecast deadlines (usually 3 h after synoptic time). Creating the early version of a late model involves several steps; as an example, we consider creating a 1200 UTC GRCI from the 0600 UTC GRCS. First, the GRCS track and intensity forecast data are interpolated to 3-hourly resolution and smoothed. Next, the smoothed GRCS data valid at 1200 UTC are compared to the TC's actual 1200 UTC position and intensity. Differences between the two represent offsets (corrections) that are then applied to the entirety of the smoothed GRCS track and intensity data. Finally, the corrected GRCS forecasts are sampled at appropriate forecast lead times to construct the 1200 UTC GRCI (e.g., the smoothed and adjusted GRCS forecast at 30 h becomes the 24-h GRCI forecast). This algorithm for generating the early version of a late dynamical model is known as the "interpolator". Note that with some operational models, the intensity offset corrections may be tapered off to zero at longer forecast leads. As will be shown below, however, the AIWP models have large and persistent intensity biases that argue for application of the offsets throughout the forecast period.

Table 1 listed the ATCF identifiers for the late and early versions of the models used in the verifications presented here. The verifications follow NHC standard procedures (e.g., Cangialosi et al. 2023). The ATCF best-track position and intensity valid at each forecast time provide the verifying data used to evaluate the AIWP forecasts. Unless otherwise indicated, all verification results shown here represent homogeneous model samples.

The track forecast error is the great-circle distance between the forecast and best-track position, and the intensity error is the difference between the forecast and best-track maximum wind. The sample mean absolute error and bias of the intensity errors are both calculated. Following Aberson and DeMaria (1994), statistical significance between model errors is evaluated using a two-sided paired-t test, applied to a head-to-head verification of the two models being compared. To account for serial correlation across forecast cycles, an effective sample size is used in the calculation of the t statistic and the degrees of freedom; forecasts separated by less than 18 h are assumed to be serially correlated. The 95% level is considered the threshold for statistical significance.

Previous evaluations have shown that TC circulations in AIWP models tend to weaken with time (e.g., Bouallegue et al. 2024). To assess this issue, a new metric was developed to measure the percentage of cases at each forecast time that the CIRA or operational TC trackers were able to detect a model vortex when the best track indicated that one should be present. This new metric, the detection rate, only includes cases for which all models being verified diagnosed a vortex at t=0; this allows a fair comparison of model detection rates across a homogeneous sample.

Out of the more than 1200 AIWP forecasts in the NH sample to be described in section 5, three cases featured spectacular intensity failures (maximum winds greater than 500 kt!) after about 96 h into the forecast. All were produced by FCN1 for two western North Pacific TCs: two for typhoon Mawar (WP022023) in late May and one for Typhoon Lan (WP072023) in early August. Examination of the FCN1 model fields for Mawar showed that a very small-scale anticyclonic vortex at 10-m appeared to the east of the TC center about 96 h into the forecast, and which amplified over the following day to contain winds exceeding 500 kt. This area of unrealistic winds spread vertically and horizontally, primarily in the east-west direction from the TC, and by 144 h covered several thousand kilometers in a narrow latitude band. In the one Lam case, a small area of high winds first appeared over central China after about 120 h and then spread similarly, eventually reaching the TC circulation by 168 h.

Because these extreme failures would have been easily identified in real time through the most cursory of quality-control checks and would likely not have been considered by forecasters, the three runaway FCN1 forecasts were removed from the sample of forecasts to be verified. The fact that no such failures occurred with FCN2, which replaced FCN1, or with either of the other tested AIWP models, illustrates the progress that has been made with AIWP models over the past two years.

## 5. Verification results

### a. Evaluation of the CIRA TC tracker

The CIRA TC tracker used to locate the TC positions and estimate intensity from the AIWP models was first evaluated to determine if the choice of tracker had any influence on the results. For that purpose, the CIRA TC tracker was applied to 0.25° GFS forecast fields (GFSS) and its errors were compared to those from the operational GFS (GFSO), which uses a more sophisticated tracker that includes multiple levels and variables. As described in the Appendix, the mean track and intensity error, intensity bias and detection rate for GFSS were nearly identical to GFSO over our northern hemisphere sample of cases. Thus, it is unlikely that the use of the CIRA tracker affected the AIWP verification results. As a reminder, in the validations below, all the operational model forecasts included for comparison were obtained from the a-decks and used the operational trackers.

### b. Late models

As a starting point, the AIWP late-model forecasts from 2023 were verified for the northern hemisphere (i.e., all AL, EP, CP, WP, and IO cases). None of the AWIP models included 2023 data in their training sample, so these cases are representative of an operational model configuration. For comparison, the late versions of the operational GFS (GFSO) and ECMWF (ECMO) deterministic forecasts were included in the homogeneous sample. Figure 1 shows the track errors, where NT is the number of homogeneous track forecasts at each forecast time (sample size). The sample size is reduced by about 50% starting at 96 h because ECMO forecasts only run to 84 h at 0600 and 1800 UTC.

Figure 1 shows that the ECMO errors were smaller than those from GFSO at all forecast times from 12 to 168 h. The AIWP models had errors comparable to each other and to ECMO, except for FCN1, which had errors similar to GFSO at most forecast times. The differences (degradations) of FCN1 relative to the other three AIWP models were statistically significant at several times between 12 and 168 h. GRCS was the best performing track model at 96 h and later, with errors up to 13% smaller than those the next best model (ECMO), although those differences were not statistically significant.

Previous studies have shown larger improvements of AIWP track forecasts relative to the ECMWF (e.g., Lam et al. 2023) than what were found here. This difference might be due to the other studies' more restrictive verification samples that excluded the depression stages, different basins and years examined, and initialization with ECMWF or ERA5 analysis fields instead of the GFS used in this study. Nevertheless, the results in Fig. 1 indicate that the AIWP models have great potential utility for TC track forecasting, based on their favorable comparison with ECMO and GFSO, two of the best-performing operational track models (e.g., Cangialosi et al. 2023).

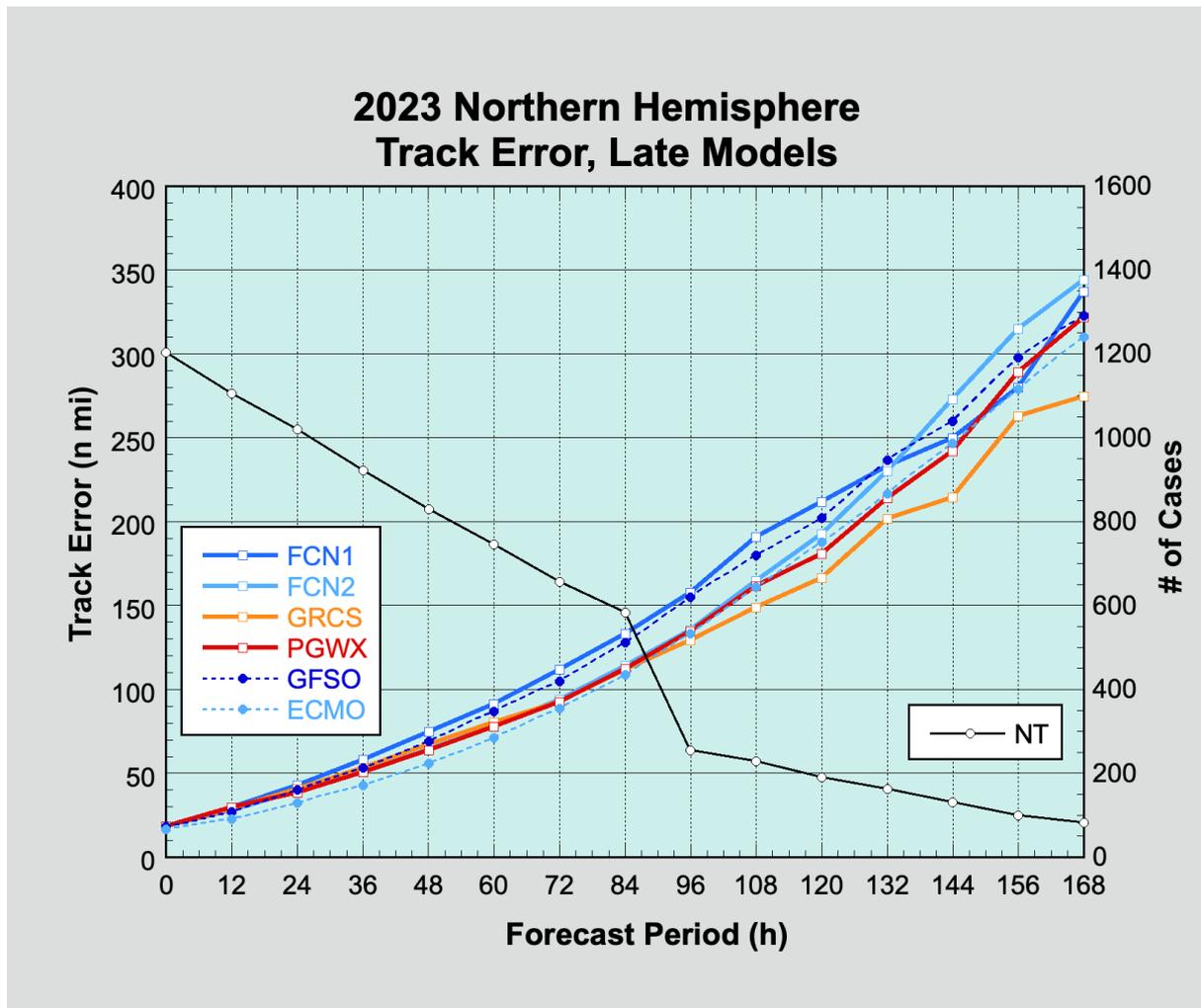

Figure 1. Mean track errors for the AIWP models and the late versions of the GFS and ECMWF models, for a homogeneous sample of northern hemisphere cases from May-Nov 2023. NT is the sample size.

Figure 2 shows the detection rate for the AIWP models, GFSO, and EMCO. The GFSO had a detection rate of almost 100% through 168 h, but the ECMO detection rate decreased to about 95% by 84 h. The large decrease in the detection rate for ECMO after 84 h is an artifact of the shorter runtimes of the 0600 and 1800 UTC ECMWF model, as mentioned earlier. The detection rates of all the AIWP models except FCN1 were comparable to the GFS through 120 h, and then a little lower after that. PGWX and GRCS had the best longer-range detection rates, which were almost as good as those for GFSO. Each of the AIWP models had a better detection rate than ECMO through 84 h; no direct comparisons are possible beyond 84 h due to the lack of 0600 and 1800 UTC ECMO forecasts.

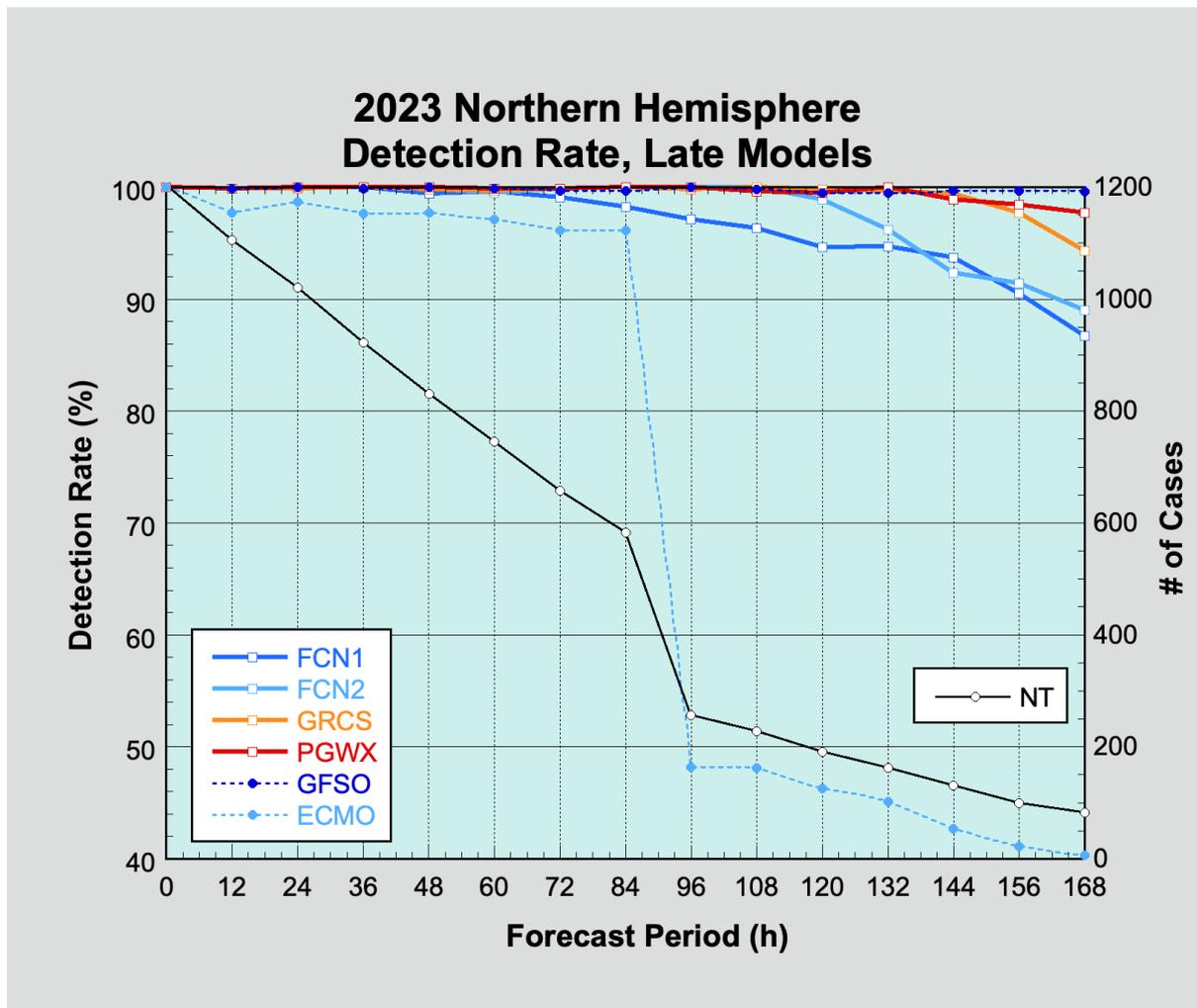

Figure 2. Same as Figure 1 but for the detection rate.

Figure 3 shows the intensity errors and biases of the AIWP models, GFSO, and ECMO. The t=0 h errors and biases for GFSO and the AIWP models are the same because the AIWP models were initialized with the GFS; the ECMO errors are larger at t=0 h. After t=0 h, the GFSO and ECMO errors grew slowly while the bias stayed roughly constant. In contrast, the AIWP errors increased rapidly after t=0 h and then leveled off at about 40 kt after 72 h. The AIWP model intensity degradations relative to the GFSO and ECMO were statistically significant from 12 to 168 h. The magnitudes of the AIWP low biases were almost the same as the mean absolute error, which indicates that nearly every intensity forecast was too low. The GFSO and ECMO intensity forecasts usually have larger errors than those from statistical-dynamical and regional dynamical hurricane models, so the fact the AIWP errors in Fig. 3 are even larger than those indicates they would have little direct utility for operational forecasting. The low intensity bias found here is consistent with other studies (e.g., Boualleque 2024).

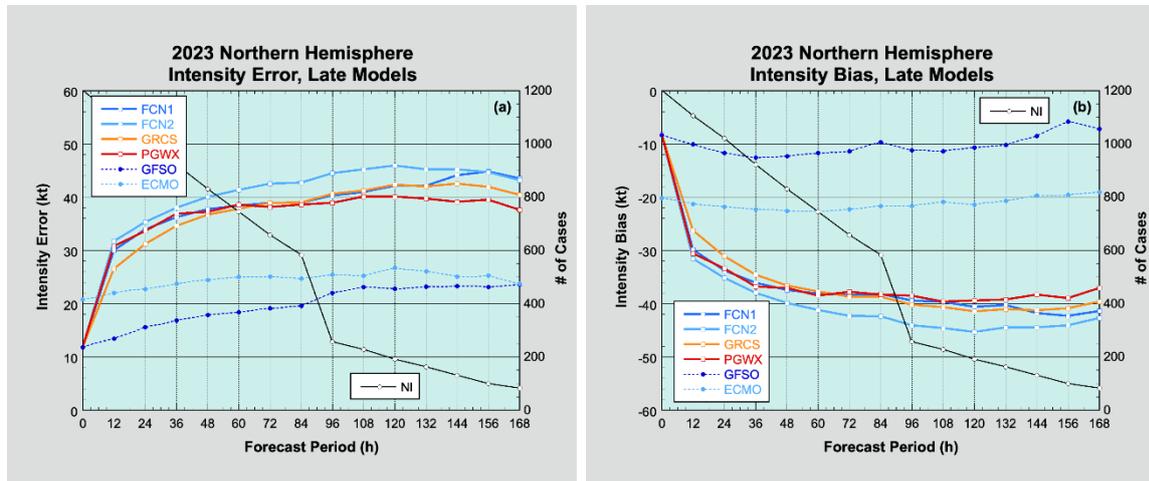

Figure 3. Intensity error (a) and bias (b) of the AIWP models and the late versions of the GFS and ECMWF models for a homogeneous sample of northern hemisphere cases from May-Nov 2023. NI is the sample size.

*c. Early models*

As noted earlier, dynamical models are generally not available in time to be used as guidance for operational TC forecasts. To better evaluate the utility of AIWP models for use in operations, early versions were created from the 6-h-old (or occasionally from 12-h-old if a 6-h-old forecast was not available) late forecasts using the standard interpolation procedure described in section 2. Errors of the generated early AIWP models were then compared with several representative early operational models. The impact of the early AIWP models on the consensus of commonly used early models was also evaluated. Because of differences in the availability of operational early models across the JTWC and NHC/CPHC basins, the early model verifications in this section were restricted to the NHC/CPHC basins (AL, EP and CP). Also, the results are only shown through 120 h because some operational models, as well as the NHC/CPHC public forecasts, are still limited to that lead time.

Figure 4 shows the early model track verifications. In addition to the early AIWP models, included are the NHC/CPHC official forecast, a climatology and persistence baseline model (TCLP, DeMaria et al. 2022), early versions of two regional hurricane models (HWFI and

File generated with AMS Word template 2.0

HMNI, Alaka and co-authors 2024, Mehra et al. 2018), and two global models (EMXI and GFSI) (see Table 1). The HAFS model (Hazelton et al. 2023), a replacement for the regional HWRF and HMON, was not available for the early part of the 2023 season and so was not included here.

Figure 4 shows that all the early models and the OFCL forecasts had much smaller track errors than the TCLP baseline and so all are considered highly skillful. The regional HWFI and HMNI errors were a little larger than the other models, especially at longer ranges. All AIWP models except FCN1 were comparable to the official forecasts and the global models through about 72 h, and then had lower errors at 96 and 120 h. The GRCS errors were the lowest of the AIWP models at 96 and 120 h and were 15 and 28% smaller than the best operational model (EXMI) at those times. Except for FCN1, the differences between the early AIWP models and the operational HWFI, HMNI, and GFSI were statistically significant at many of the times between 36 and 120 h. These results suggest the AIWP track forecasts have considerable potential as guidance for operational forecasts, especially at the longer forecast lead times.

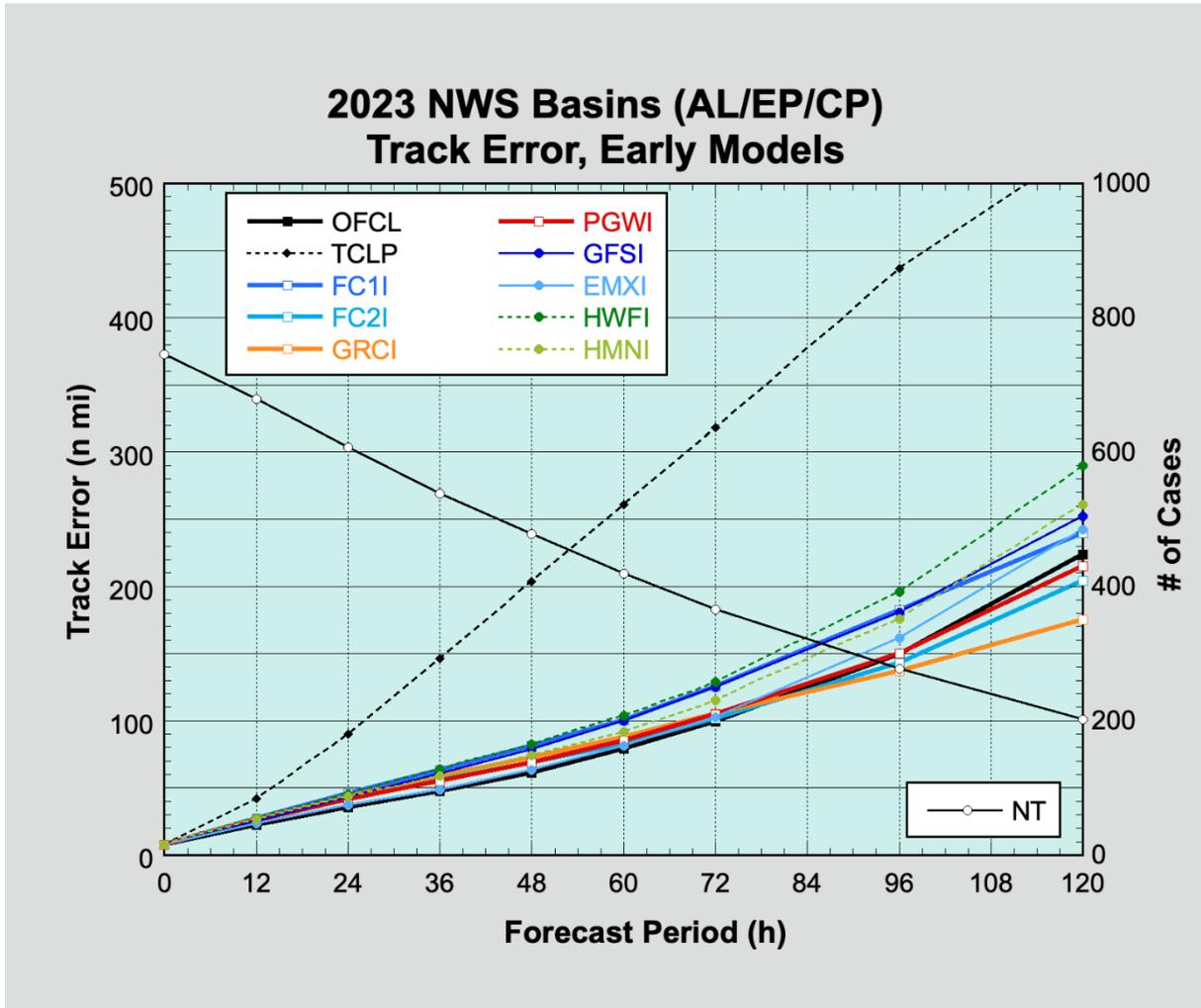

Figure 4. Track errors of the early versions of the AIWP models, the NHC/CPHC official forecasts, and five early operational track models.

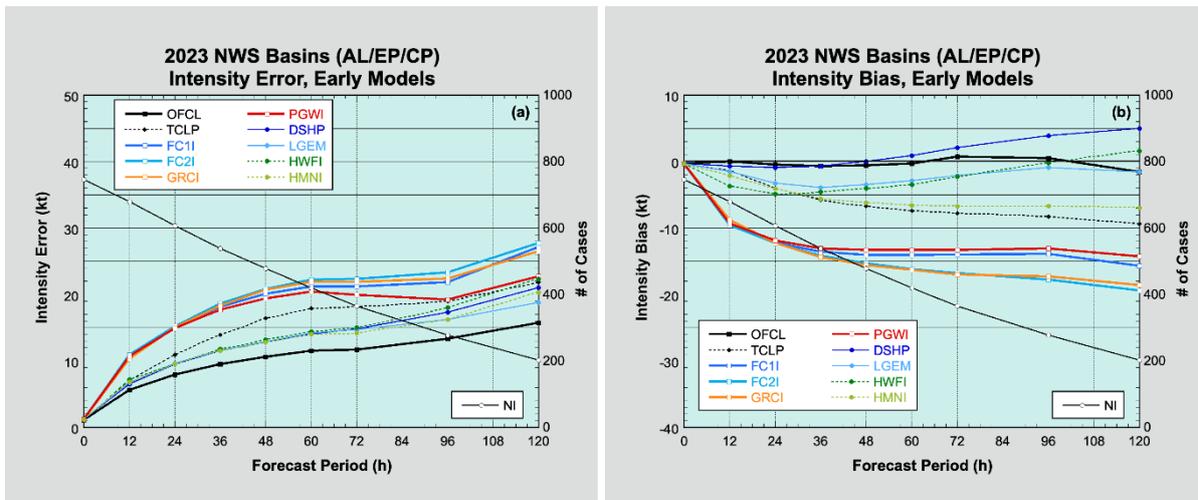

Figure 5. Intensity errors (a) and biases (b) of the early versions of the AIWP models and five operational track models.

Figure 5 shows the errors and biases of the early model intensity forecasts. The operational intensity models shown for comparison comprise two statistical dynamical models (DSHP and LGEM), the early versions of the two regional hurricane models (HWFI and HMNI), and the skill baseline TCLP. Comparing Fig. 5a with Fig. 3a shows that the interpolator significantly reduced the AIWP intensity errors. Although the samples in Fig. 5a and 3a are not the same, the AIWP models all had a consistent low bias, which is partially corrected by the interpolator due to the difference between the 6-h intensity forecast and the current intensity. PGWI was the best early AIWP model, and its differences from the other early AIWP models were statistically significant from 72-120 h. However, the early AIWP intensity errors were still larger than those of the baseline TCLP, indicating that they have no forecast skill. The AIWP intensity errors were also much larger than the errors of OFCL and the four operational models. Figure 5b shows that, despite the adjustments made by the interpolator, a low bias is still a large contributor to the AIWP early version intensity errors.

*d. Examination of AIWP model intensity bias*

Figure 3 shows that the AIWP model intensity errors increase very rapidly in the first few days of the forecast, due mainly to the rapid increase in the magnitude of the low bias. There are several possible reasons for this behavior, including the training data and the choice of the performance metric (loss function) that was optimized.

(1) Training data

All the AIWP models evaluated in this study were trained on ERA5 reanalysis fields. The maximum wind in the ERA5 has a low bias (Dulac et al. 2023) as does the ECMWF forecast in Fig. 3b, which has a bias of about -20 kt at t=0 h. However, the magnitude of the AIWP intensity bias in Fig. 3b exceeds that of ECMO after just 12 h and continues to increase through about 72 h, which suggests that the training data is not the only reason for the low bias.

(2) Performance metric

Bonavita (2024) examined the representation of sub-synoptic and mesoscale phenomena in the FCN2, PGWX, and GRCS models, and showed that those AIWP models tended to "blur" smaller scale features and reduce their amplitudes, similar to the results for the TC intensity forecasts in Fig 3. A spectral analysis of PGWX showed that the model did not properly capture scales of motion below 500-700 km, and the amplitudes of those scales in the forecast fields became much lower than those in the ERA5 reanalyses that were used to train the model. They suggested that the damping and blurring of the smaller scales may result from training the models to minimize the mean-square or absolute error of the forecast fields.

To further investigate the impact of minimizing mean square errors on the AIWP low intensity bias, a simple experiment was performed that minimizes the root mean square difference between wind fields of an idealized vortex and the same vortex displaced by a distance equal to the AIWP model track error. As shown in Fig. 1, the AIWP average track errors increase just a little faster than linearly through seven days. The 120-h track error of the AI models is ~160 n mi, or about an increase of 8 n mi (15 km) every 6 h. The TCs in the model fields are localized vortices with maximum winds very close to the center. With an incorrect location, the least squares solution in a 6-h forecast would probably be minimized by lowering the winds near the TC center and possibly increasing them away from the center to reduce the double-error penalty caused by the displacement, leading to a blurring of the wind field around the TC.

To quantify the impact of minimizing the wind field differences, a symmetric Rankine vortex given by

$$V = V_m(r/r_m) \qquad r \leq r_m \qquad (1a)$$

$$V = V_m(r_m/r)^x \qquad r > r_m \qquad (1b)$$

was used, where $r$ is the distance from the TC center, $V_m$ is the maximum wind, $r_m$ is the radius of maximum wind and $x$ is a shape parameter. Equation (1) was used to calculate a horizontal wind field on a 20-km grid, which is close to the grid spacing of the AIWP models, and then recalculate the wind field for the vortex with the center displaced by 15 km to represent a 6-h track error. The initial values of $r_m$ and $x$ were set to 40 km and 0.5, respectively, which are representative of a typical TC. The initial value of $V_m$ was set to 53.3 kt, to match the average initial intensity of the AIWP models in the late model verification sample. After the displacement, the values of $V_m$ and $r_m$ were determined to minimize the least-squares difference between the two vortices within 200 km of the original TC center.

This process was repeated 28 times to represent a 168-h forecast with a 6-h time step. The vortex with the new $V_m$ and $r_m$ values was recentered every 6 h for this calculation.

Figure 6 shows $V_m$ for the displacement experiment, the average intensity from the AIWP model forecasts, and the corresponding average best track intensity for the late model sample. The average best- track intensity shows slow intensification through ~72 h followed by a near-constant intensity and then a slow weakening by168 h. In contrast, the AIWP forecasts show rapid weakening for the first ~24 h, consistent with Fig. 3b. The intensity of the idealized Rankine vortex also decays due to a displacement that represents track error, although not as rapidly as that of the AIWP models. The Rankine vortex decay rate begins to level off later in the forecast, with a maximum wind value similar to that of the AIWP models. The value of $r_m$ that minimized the error of the displaced vortex increased by about a factor of two by 120 h as the maximum wind decreased. Thus, the idealized vortex weakened and spread out due to the displacement, consistent with the behavior of the AIWP model intensity evolution. The very rapid AIWP model intensity decay in the first 12 h in Fig. 6 is probably due to a combination of the training on ERA5 analysis and the double-error penalty.

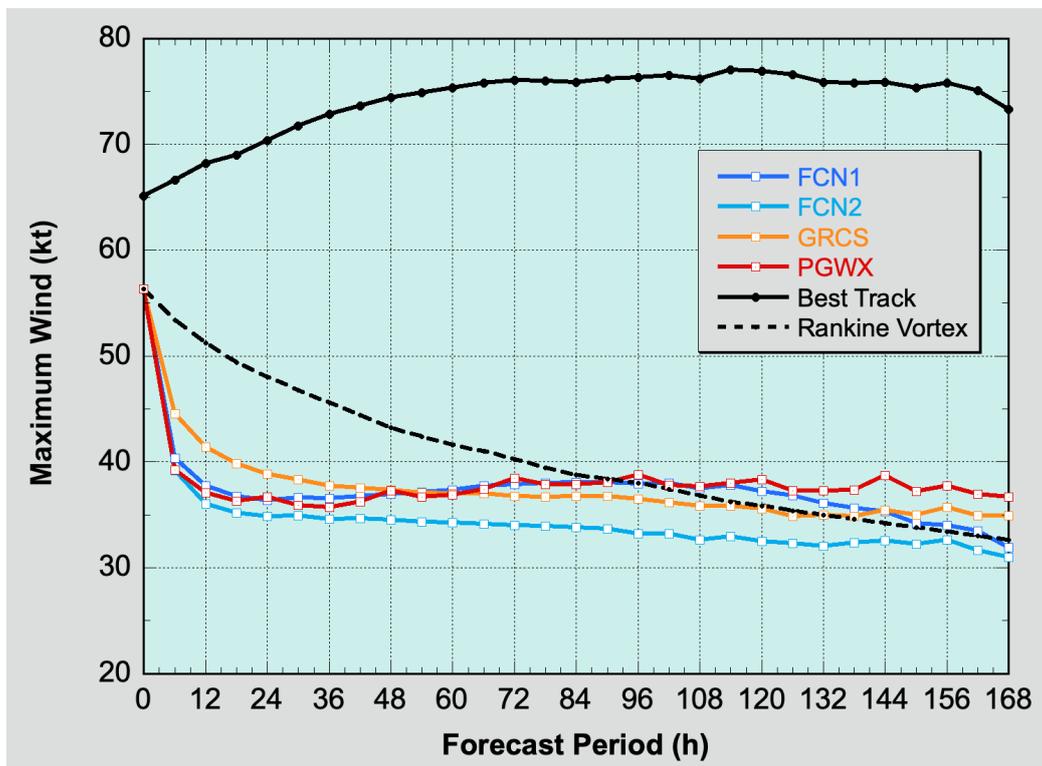

Figure 6. The evolution of the maximum wind from the Rankine vortex with a maximum wind and radius of maximum wind chosen to minimize the difference between the original

vortex and one that was displaced by 15 km every 6 h. Also shown are the average intensities of each of the four AIWP model forecasts and best track for the late model sample.

As an example of the vortex evolution in the AIWP models, Fig. 7 shows the azimuthally averaged 850-hPa tangential wind from PGWX forecasts for Hurricane Dora (EP052023) and Hurricane Lee (AL132023), as calculated by the CIRA TC tracker. The best-track intensity for Dora was a constant 115 kt during this 24-h period as the TC moved south of the Hawaiian Islands near 12°N. The best-track intensity for Lee during the 24-h period was nearly constant (between 100 to 105 kt) as it moved west-northwestward near 23°N to the north of Puerto Rico. Despite the nearly steady-state maximum winds in the best tracks, the inner core winds in both cases rapidly decrease, and the radius of maximum wind expands during this 24-h period, qualitatively similar to the results from the idealized displaced Rankine vortex. This result suggests that modifications to AIWP model training, such as using a customized loss function, will be needed for accurate intensity forecasting.

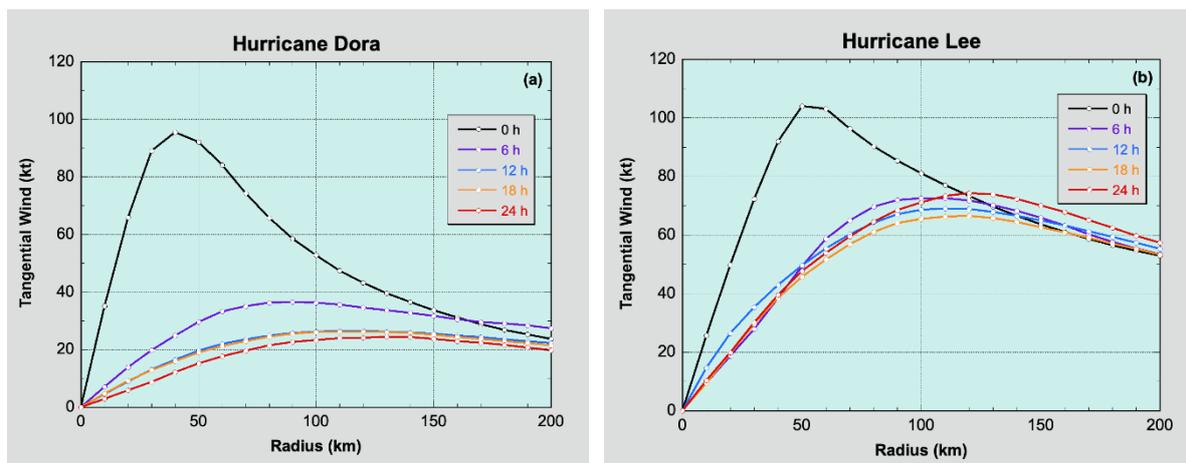

Figure 7. The azimuthally averaged 850-hPa tangential wind at 0-24 h for PGWX forecasts for (a) Hurricane Dora initialized at 1200 UTC 07 August 2023, and (b) Hurricane Lee initialized at 0000 UTC 11 September 2023.

## 6. AIWP impact on consensus forecasts and model spread

The most accurate operational track and intensity models are usually consensus models, which are combinations (generally simple averages) of the forecasts from several high-performing individual models. The consensus approach can be particularly effective when the member models are independent of each other, i.e., their errors arise from different sources. AIWP models offer a distinct potential here, because they are by nature very different from NWP models.

To determine if the AIWP models could contribute positively to the consensus, they were added one at a time and then all at once to one of the primary consensus models run by NHC and CPHC in 2023. In addition, the spread of the track and intensity AIWP forecasts was calculated to evaluate the possible application of AIWP models to ensemble forecast systems.

a. *Consensus forecasts*

NHC generates several combinations of track and intensity consensus models each season, evaluating their composition on an annual basis (Cangialosi 2023). To evaluate the contributions of the AIWP models, we used here the same base consensus that NHC recently used to test the contributions of HAFS to the consensus (Cangialosi, personal communication). This base consensus for track comprises the early global and regional track models GFSI, EGRI, HWFI, CTCI, EMNI, and HMNI, while the base intensity consensus comprises the early statistical and regional models DSHP, LGEM, HWFI, CTCI, and HMNI (Table 1).

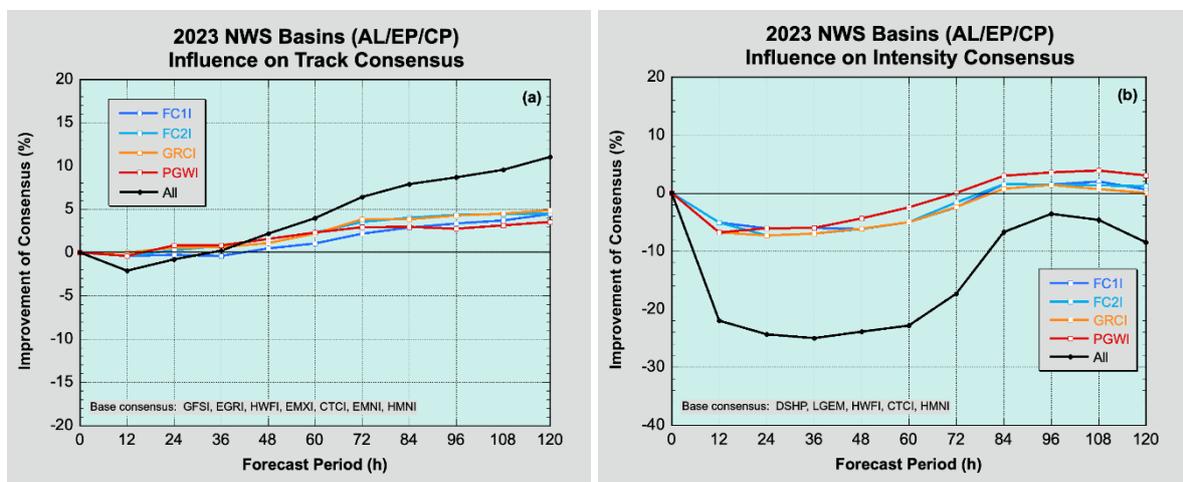

Figure 8. The percent improvement of the consensus forecasts when the AIWP model forecasts were added to the base track consensus (a) and intensity consensus (b). Note that the various lines in this figure do not represent homogeneous samples.

Figure 8a shows the impact of the early AIWP forecasts on the base consensus when they were added one at a time and all at once. Each AIWP model individually improved the track consensus, by up to 5% at the longer ranges. When all four were added the track consensus improved by 11% at 120 h, with improvements that were statistically significant at 72-120 h. Some of the improvements of the individual AIWP models (except for FCN1) were statistically significant at multiple times between 60 and 120 h. This result suggests that the AIWP models have considerable potential to improve operational track forecasts through their inclusion in consensus models.

22File generated with AMS Word template 2.0

Figure 8b shows the impact of the early AIWP models on the intensity consensus. When all four were added, the consensus was degraded at all forecast times, and the degradation was statistically significant at 12-60 h. Somewhat surprisingly, the AIWP models slightly improved the intensity consensus at some of the later forecast periods, although none of these improvements were statistically significant. The procedure used to generate the early AIWP intensity forecasts greatly reduced the magnitude of the low intensity bias of the late models (compare Fig. 3b with 5b), but the early forecasts still had a substantial low bias. Nevertheless, the fact that a simple bias correction can produce a near-neutral impact on the consensus leads to some optimism that more sophisticated statistical post-processing might be able to extract useful intensity forecast information from the AIWP models.

b. *Model spread*

The computational efficiency of the AIWP models holds potential for producing very large forecast ensembles. Current ensemble systems use a single model with perturbed initial conditions, forecasts from different models, and/or forecasts from the same modeling system with perturbed physical processes. As a first step in evaluating the ensemble applicability of AIWP models, the track and intensity spread of the four AIWP models was calculated, with the spread defined as the mean distance of each model's forecast from the ensemble mean. For comparison, the same calculation was performed with representative operational track and intensity models: EMXI, GFSI, HWFI, and HMNI for track, and DSHP, LGEM, HWFI, and HMNI for intensity.

Figure 9 shows that the track spread of the AIWP models was only a little smaller than that of the four-model reference ensemble. However, the intensity spread of the AIWP models was less than half that of the reference ensemble. This result is consistent with Selz and Craig (2023), who showed that error growth in AIWP can be comparable to that of physically based models on larger scales, but much slower than physically based models on smaller scales. Because TC tracks are largely controlled by synoptic-scale steering, while intensity changes are also strongly affected by smaller-scale physical processes such as air-sea energy exchanges and moist convection, we are more optimistic about the applicability of AIWP ensembles to track forecasts than to intensity forecasts.

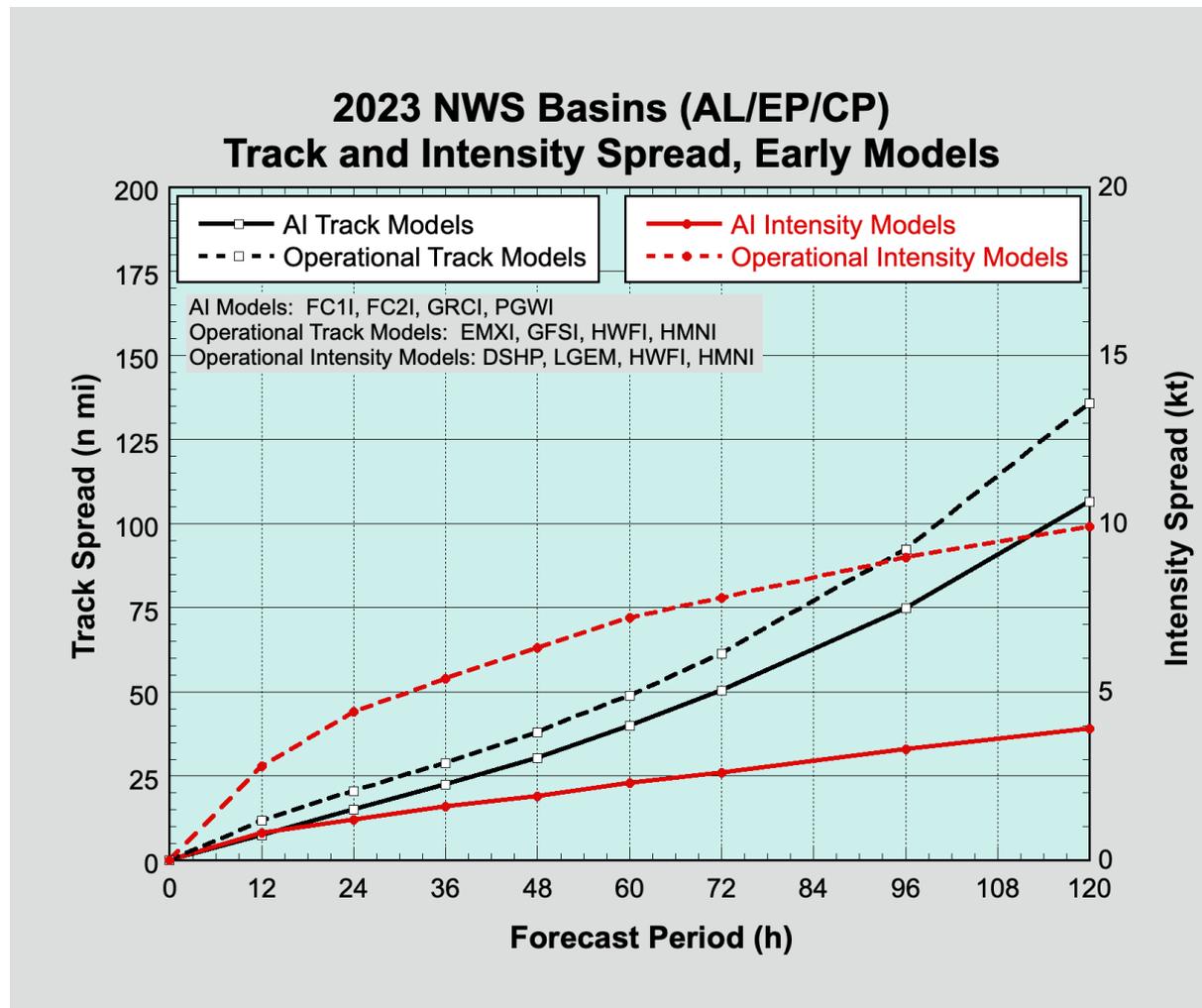

Figure 9. The track and intensity spread from the four AIWP models and representative sets of four operational track and intensity models.

## 7. Concluding remarks

Seven-day TC track and intensity forecasts from four open-source AIWP models (FCN1, FCN2, GRCS, and PGWX) initialized with 0.25º GFS analyses were verified against best-track data from NHC, CPHC, and JTWC for all northern hemisphere cases for the period May-Nov 2023. The sample included more than 1200 forecasts from the AL, EP, CP, WP and IO basins. Operational forecasts from the GFS and ECMWF models were included for

File generated with AMS Word template 2.0

comparison. The NHC verification software package and procedures were used to calculate all the forecast errors for this study.

Results showed that FCN2, GRCS, and PGWX track forecasts had errors that were comparable to or smaller than those from the GFS and ECMWF, which are typically the best-performing individual track models. GRCS was the best-performing model at the later forecast times. The track errors from FCN1 were a little larger, but still comparable to those from the GFS.

A new metric called the detection rate was added to the NHC verification package; this measures the percentage of times a forecast from a model was available when there was a verifying position in the best track. The operational GFS had the best detection rate, with values very close to 100% through 168 h. PGWX had a slightly lower detection rate, but was still above 98% through 168 h. GRCS, FCN1, and FCN2 also had very good detection rates, which were nearly all above 90% through 168 h. Detection rate comparison with the ECMWF was not possible for lead times of 96 h and longer due to the lack of 0600 and 1800 UTC ECMWF forecasts at those lead times. However, the AIWP model detection rates were greater than those of the ECWMF at 12-84 h.

Dynamical models are usually not available in time to be used for operational forecasts, and so are called late models. AIWP models also fit into the category of late models, as their initialization relies on dynamical model analyses. A standard procedure used by operational forecast centers was applied to create early versions of the AIWP model forecasts. The early AIWP track errors were comparable to those of the official forecasts and best-performing operational track models. The impact of including the early AIWP models in consensus track forecasts was also evaluated, and results showed that the consensus was improved by up to 11% at the longer ranges. To put that improvement in perspective, the NHC official Atlantic five-day track forecast error improved by about 2% per year between 2001 and 2023 (https://www.nhc.noaa.gov/verification/verify5.shtml ), so an 11% represents more than five years of track error reduction. The spread of the four AIWP track forecasts was compared to the spread of four representative operational track models to help assess the utility of AIWP in ensemble forecasting. The AIWP model track spread was only a little smaller than the spread of the operational models. This result suggests AIWP models will have utility in estimating track forecast uncertainty.

In contrast to the very positive track forecast results, the AIWP intensity forecasts had very large errors compared with official intensity forecasts and the operational models – degradations that were statistically significant at all forecast times from 12-168 h. The large intensity errors were related to a substantial low bias that increased rapidly during the first 48 h of the forecasts. The early version of PGWX had lower intensity errors than the other early AIWP models, but its errors were still larger than those from TCLP, indicating no skill. The interpolation procedure used to create the early models partially corrects the low bias, leading to a near-neutral result when the AIWP models were added individually to the intensity consensus. The spread of the intensity forecasts from the early AIWP models was much smaller than the spread from four representative operational intensity models, casting doubt on their immediate utility for ensemble applications.

The rapid weakening seen in AIWP model intensity forecasts might partially be due to training on ERA5 reanalysis, which themselves have low bias in the maximum winds. Another contributing factor appears to be an interaction between the track errors and vortex structure when minimizing RMS errors. Experiments with an idealized Rankine vortex displaced by a distance roughly equal to the average 6-h track error of the AIWP models showed that the AIWP model training process itself may preferentially favor weakening vortices. It appears that different model training data and procedures, such as customized loss functions, might be needed to improve the AIWP TC intensity forecasts.

The small track errors and large intensity errors from the late AIWP models were consistent with results from previous studies. This study extended previous work by evaluating AIWP forecasts in the way they would be used operationally, using early versions of the models and following NHC operational verification standards. The small track error, high detection rates, positive contributions to consensus track forecasts, and the error spread results indicate that the AIWP models have great potential for improving operational track forecasts. Further improvements might be made by including AIWP track forecasts in corrected consensus models (Simon et al. 2018) that unequally weight the contributions from individual models based on past performance and other factors.

The large intensity errors, substantial low biases and small error spread of the AIWP forecasts indicate that the AIWP models currently have little direct applicability to operational intensity forecasting. It is possible that more sophisticated bias-correction methods might make the AIWP intensity forecasts useful, especially for the longer ranges when the spurious weakening diminishes. However, significant changes to AIWP models will likely be needed for the intensity forecasts to show the same operational potential as the track forecasts.

*Acknowledgments.*

This research was partially supported by the Hurricane Forecast Improvement Program (HFIP), part of the CIRA CA grant NA19OAR4320073. We would like to thank CIRA and NOAA Global Systems Laboratory for producing and providing the AI models reforecast data archive.

*Data Availability Statement.*

The NHC and CPHC ATCF a-decks (official and model forecasts) and b-decks (best tracks) are available from ftp://ftp.nhc.noaa.gov/atcf and the JTWC ATCF b-decks are available from https://www.metoc.navy.mil/jtwc/jtwc.html. The JTWC ATCF a-decks are not publicly available and were used by permission from JTWC for this research under the condition that they are not redistributed. The AIWP model fields are available from https://noaa-oar-mlwp-data.s3.amazonaws.com/index.html .

APPENDIX

**The CIRA TC Tracker**

The center-finding method applied to the AIWP models is based on the vortex removal procedure from the SHIPS model (DeMaria 2010). This method was used because it only requires horizontal winds from one level (850 hPa) to track the center so it can be applied to all the AIWP models used in this study, including FCN1, which has the most limited number of fields. Because the results depend on the properties of the TC tracker, which have not been documented elsewhere, and the tracker parameters differ from those in the SHIPS model, the details of the CIRA TC tracker are described here.

Table A1 summarizes the procedures of the CIRA TC tracker. Several of the steps require conversions between lat/lon increments and distances. A simple tangent-plane approximation centered on the first guess center position at each forecast time is used for the conversions. Linear interpolation is used to obtain the u and v horizontal wind components on the cylindrical grid in step 2. Steps 3-4 differ from the single-level tracker described by Marchok (2021) in that the averaging of the tangential wind is not an area average. Instead, a simple radial average is used to weight the tangential wind near the TC center more when finding the center position that maximizes the circulation.

1. Use the current lat/lon of all active TCs from the ATCF a-decks as the first guess positions.

2. Interpolate the t=0 h model 850 hPa horizontal wind components (u, v) to a cylindrical grid with radial grid spacing $\Delta r$ and azimuthal spacing $\Delta \alpha$ centered on the ATCF TC position.

3. Calculate the tangential wind from u and v on the cylindrical grid and perform an azimuthal average from r=0 to $R_{max}$.

4. Radially average the azimuthally averaged tangential wind from r=0 to $R_{max}$.

5. Perturb the first guess lat/lon by ±N increments of a distance $\delta s$ in the east-west and north-south directions and repeat steps 2-4.

| |
|---|
| 6. Find the location from the displacements in step 5 that maximizes the radially and azimuthally averaged tangential wind (circulation maximum) and use that as the updated center. |
| 7. Reduce the displacement distance δs by a factor F and repeat steps 5 and 6. |
| 8. Repeat step 7 M times and use the location after the last iteration as the TC center position. |
| 9. Search the 10-m wind speed on the original AIWP model grid with a distance ±$D_i$ from the center position in step 8 and use that for the TC intensity. |
| 10. Check to see if the TC is strong enough to track. There are two conditions, where the intensity needs to be greater than a specified wind speed ($V_{min}$) and/or the average tangential wind needs to exceed a specified value ($V_{t\_min}$). If the conditions are not met, the tracker stops. |
| 11. Extrapolate the center position from step 8 ahead to the next model time step for the new first guess. For the extrapolation at t=0 h, use the TC motion vector from the ATCF a-deck. For all later times, extrapolate forward using the TC tracker positions from the two previous time steps. Repeat steps 2-9 to find the TC center and intensity at each time step to the end of the forecast. |

Table A1. The steps in CIRA TC tracker used in the AIWP verification study.

The CIRA TC tracker in Table A1 requires the specification of several parameters. These were modified from those used in the SHIPS model based on a comparison between positions from the CIRA TC tracker applied to GFS 0.25º model output and the more general GFS operational tracker positions obtained from the ATCF. Table A2 shows the values of the tracker parameters used in this study. With these values, the CIRA TC tracker almost exactly reproduced the mean track errors, detection rates, mean intensity errors, and intensity biases of operational GFS tracker results from the ATCF a-decks for the 2023 northern hemisphere sample.

| Parameter(s) | Description | AIWP tracker value(s) |
|---|---|---|
| $\Delta r$, $\Delta \alpha$ | Radial and azimuthal spacing of the cylindrical grid used to average the tangential wind | 10 km, 22.5º |
| $R_{max}$ | Radial distance to average the tangential wind | 300 km |
| $\Delta s$ | Distance to perturb the first guess TC position | 40 km |
| N | Number of increments of δs to perturb the first guess TC position | 4 |
| F | The factor to reduce δs in each iteration | 0.6 |
| M | Number of times to reduce δs and refine the center position | 7 |
| $D_i$ | Search distance for the maximum 10 m wind | 300 km |
| $V_{min}$, $V_{t\_min}$ | Minimum values of the intensity and averaged tangential wind needed to continue the tracker | 10 kt, 2 m/s |

Table A2. Parameters used in the CIRA TC tracker for the AIWP models.

File generated with AMS Word template 2.0

File generated with AMS Word template 2.0